# Deep Learning For Classification Of Chest X-Ray Images (Covid 19)

Benbakreti Samir[1], Said Mwanahija[1], Benbakreti Soumia[2], Umut Özkaya[3]

[1] Specialty Department, Ecole Nationale des Télécommunications et des Technologies de l'Information et de la Communication (ENSTTIC), Oran, Algeria.

[2] Laboratoire des Mathématiques, University of Djillali Liabes, Sidi Bel Abbes, Algeria.

[3] Electrical ans Electronic Engineering, Konya Technical University, Turkey

## ABSTRACT

In medical practice, the contribution of information technology can be considerable. Most of these practices include the images that medical assistance uses to identify different pathologies of the human body. One of them is X-ray images which cover much of our work in this paper. Chest x-rays have played an important role in Covid 19 identification and diagnosis. The Covid 19 virus has been declared a global pandemic since 2020 after the first case found in Wuhan China in December 2019. Our goal in this project is to be able to classify different chest X-ray images containing Covid 19, viral pneumonia, lung opacity and normal images. We used CNN architecture and different pre-trained models. The best result is obtained by the use of the ResNet 18 architecture with 94.1% accuracy. We also note that The GPU execution time is optimal in the case of AlexNet but what requires our attention is that the pretrained models converge much faster than the CNN. The time saving is very considerable.

With these results not only will solve the diagnosis time for patients, but will provide an interesting tool for practitioners, thus helping them in times of strong pandemic in particular.

## 1. Introduction

Computerized Tomography (CT) and X-ray scans are frequently used for chest imaging. An X-ray is a scan of the body that looks for pneumonia, tumors, fractures, and lung infections. An upgraded X-ray machine called a CT scan can produce sharper images of bones, tissue, and organs. Compared to CT, the X-ray approach is simpler, faster, and more affordable, but it is also more dangerous. Doctors can visually diagnose viral bacterial infections, viruses like covid 19 [1], and other infections by examining chest X-ray images. The technique of visual diagnosis is typically unappealing, time-consuming, and inaccurate, because it can result in low accuracy and requires specialized human resources.

Coronavirus disease 2019 (COVID-19) is an infectious disease brought on by the coronavirus strain known as severe acute respiratory syndrome coronavirus-2 (SARS-CoV-2) [2]. It is a lung infection that is respiratory in nature. The root of the coronavirus word is Greek (κορώνη) which means "crown or halo." It relates to the virus's appearance under an electron microscope, which resembles a royal crown. Because of this, coronavirus is also known as the crowned virus. The purpose of this paper is therefore to provide a decision making tool that will lighten the burden on medical staff, especially during pandemic peaks.

## 2. Related work

Since Corona was announced as a pandemic, different projects were carried out since 2020 to 2022 that it became among an interesting subject to learn from, some of the works related to image classification for different reasons are discussed in this paragraph.

The COVID-CT dataset of 2560 images was the database used in [3], 2214 of which were used for training and the remaining 246 for testing. By employing WOA to optimize the network's hyperparameters, the model used to train ResNet-50 became the WOANet model. This last experiment looked at the accuracy of the classification using the suggested method on 246 CT scans, and found that 98.37% of them were categorized as COVID-19, while 99.18% were identified as non-COVID-19. The radiologists will be greatly assisted by this proposed WOANet in reducing the burden on the healthcare system and hospitals.

In this study [4], patients' X-ray images are used to classify patients using CNN deep learning. One of the most powerful algorithms with generative and deterministic capabilities is the capsule network (CapsNet). However, compared to the basic CNN structures, this network has been relatively more responsive to images. The dataset utilized was the NIH complete Chest X-rays [5] collection. VDSNet has a validation accuracy value of 73%, which is higher than the sample dataset's score of 70.8%.

Using a dataset of 6432 images, the DLH COVID [6] model is distinct, trustworthy, and independently created without any input from the transfer learning approach. The experimental findings from the prospective validation phase suggest that the DLH MODEL outperformed the majority of the pre-trained models since it distinguished COVID-19, pneumonia, and healthy/unhealthy patients from the image dataset with a promising accuracy of 96%.

## 3. Dataset

As seen in figure 1, a database of chest X-ray images for COVID-19 positive cases as well as images of normal and viral pneumonia was created in collaboration with medical professionals by a group of researchers from Qatar University, Doha, Qatar, and the University of Dhaka, Bangladesh, as well as their collaborators from Pakistan and Malaysia. This dataset contains 3616 COVID-19 positive cases, 10,192 Normal, 6012 Lung Opacity (Non-COVID lung infection), and 1345 Viral Pneumonia images. The COVID-19 x-ray image database was created using different sources [7, 8, 9].

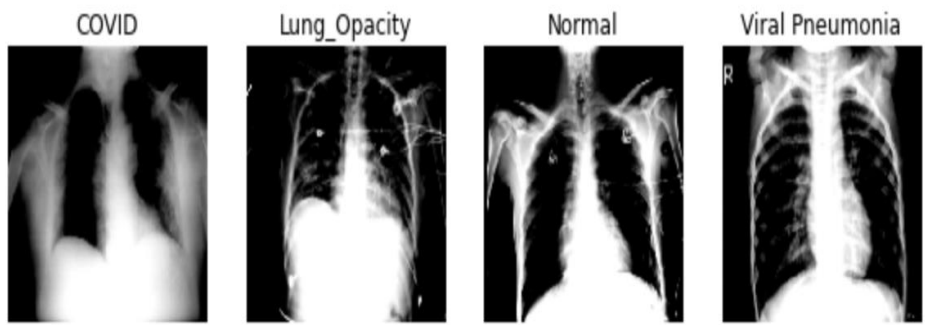

Fig. 1 CXR scans with four categories of pathology.

For training of datasets, we employed Matlab 2021 installed on computer with 64-bit operating system, windows 10 Pro, 24 GB of Random Access Memory (RAM), with an Intel(R) Xeon(R) CPU E5-2620 v3 @ 2.40GHz and Graphical Processing Unit (GPU). Eighty percent of the datasets are used for training and 20% for testing (evaluating the model performance).

## 4. Proposed Model

In this study, we analyzed the different techniques for image classification of COVID-19 using X-Ray radiographic images of the chest, then examined CNN's architecture that is based on research on the visual cortex of the cat by Hubel and Wisiel [10], and different pre-trained models: AlexNet, ResNet18 and GoogleNet in order to see the variation of answers in our work.

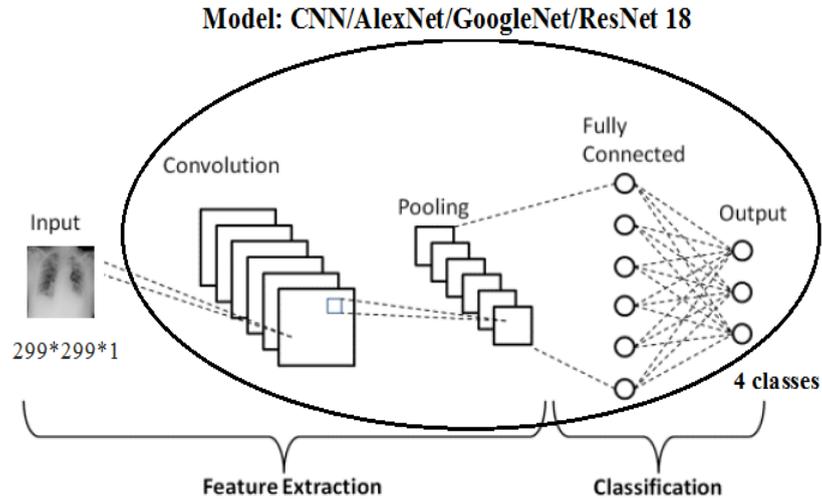

Fig. 2 The proposed model with different algorithms.

The use of deep learning methods not only allows us to process a very large number of images but at the same time allows us to skip the feature extraction step, such a cumbersome step because it is done by hand crucially.

Due to their capacity to extract features (see figure 2) and learn to distinguish between various classes, convolutional neural networks (CNNs) are the top DL tool that are widely employed in several fields of the healthcare system (i.e., positive and negative, infected). Transfer learning (TL) has made it simpler to quickly and accurately retrain neural networks on chosen datasets.

## 5. Experiments and Results

### 5.1 Experiment 1: Application of the CNN model

The structure of our CNN includes a number of layers, as shown in table I. CNN receives a CXR image with a size of 299 by 299 pixels as its input, and the rest of the architecture is mentioned in the table I.

Table 1 The architecture of the CNN model.

| Name | Type | Description of output size |
|---|---|---|
| Input layer | Input data | 299*299 |
| Conv 1 | Convolution +ReLU | 32*32*8 |
| S1 | Max pooling | 3,2 |
| Conv 2 | Convolution +ReLU | 64*64*3 |
| S2 | Max pooling | 3,2 |
| Conv 3 | Convolution +ReLU | 128*128*5 |
| S3 | Max pooling | 3,2 |
| Conv 4 | Convolution +ReLU | 256*256*5 |
| S4 | Max pooling | 3,2 |
| Conv 5 | Convolution +ReLU | 512*512*5 |
| S5 | Max pooling | 3,2 |
| Conv 6 | Convolution +ReLU | 1024*1024*5 |
| S6 | Max pooling | 3,2 |
| Fc | Fully connected | 1 Fc (4) |

The trained parameters used in this model are in the options side where all the hyperparameters used were defined including the number of epochs used (1 or 5), the mini batch (64), the learning rate is 0.001 and frequency validation is 20. The given CNN was trained using different parameters to test the accuracy for this model. We utilized the accuracy parameter to evaluate how well the trained models performed. The percentage of correctly classified images over all the images is what is referred to as accuracy. The following formula is utilized:

$$Accuracy = \frac{TP + TN}{TP + TN + FP + FN}$$

Table 2 The results of the CNN model.

| The CNN model | | |
|---|---|---|
| | 1 epoch | 5 epochs |
| Accuracy | 75.61% | 89.13% |
| Time GPU execution | 146 min 58s | 703 min 16s |

The first top accuracy after training the model using one epoch provided us with 75.61% accuracy for our 4 classes classification. As training the model using only one epoch did not provide the best result, we had to increase the number of epochs and see the performance of our model and the results for our model gave us 89.13%, which is a lot better compared to our first experiment with one epoch.

**5.2 Experiment 2: Application of the pretrained models**

**AlexNet:** With 5 convolutional layers and convolutional filter sizes of 3*3 and 2*2 for max pooling operation, AlexNet is an 8-layer convolutional neural network [11]. Fully connected layers are the final three layers. The AlexNet model's standard input size is 227*227*3.

**GoogleNet (Inception v3):** A convolutional neural network with 50 layers in depth is called GoogleNet [12]. The program, titled "Going deeper with convolutions," was developed and taught by Google. Up to 1000 objects can be classified using the pre-trained Inceptionv3 model with the ImageNet dataset [13] weights. This network's image input size was 299x299 pixels.

**ResNet18:** A convolutional neural network with 18 layers in depth is called ResNet18. Deep Residual Learning for Image Recognition, as it is known, was developed and trained by Microsoft in 2015 [14]. To address the issue of vanishing gradient that may affect the weightage change in neural networks, ResNet architectures introduced the use of residual layers and skip connections. This made training easier and allowed neural networks to get much deeper with greater performance. The network was trained on colored images with a resolution of 224x224 pixels.

In addition to the accuracy parameters, we estimated the time GPU execution for each model. The results obtained are shown in Table 3.

Table 3 The results of the pretrained models.

|  | Pretrained models | | |
| --- | --- | --- | --- |
|  | AlexNet | GoogleNet | ResNet18 |
| Accuracy | 89.93% | 91.87% | 94.1% |
| Time GPU execution | 14 min 58s | 41 min 34s | 33 min 13s |

Confusion matrix is the common approach used for evaluation of model performance based on true positive (TP), true negative (TN), false positive (FP), and false negative (FN).

The figure 3 represents the confusion matrix of the Resnet 18 model which gave the best result in terms of accuracy.

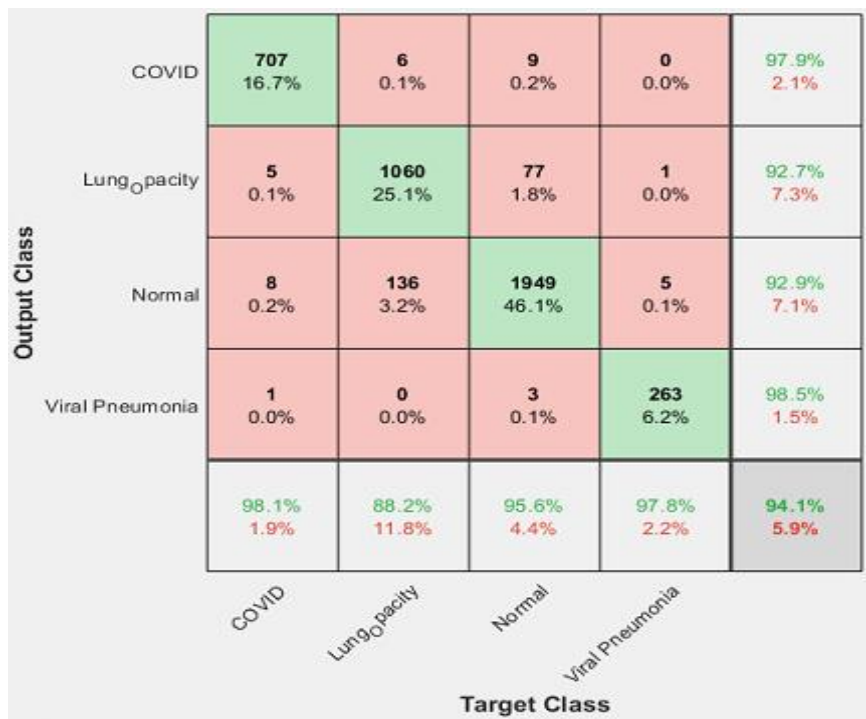

Fig. 3 The confusion matrix for the RestNet 18 model with the best result.

When the same dataset was used using the same hyper parameters the accuracy found was 89.93%, 91.87% and 94.1 % for AlexNet, GoogleNet and ResNet18 respectively. Note that the pretrained models used only one epoch. We see that the results can be improved by using pretrained architectures, attaining an accuracy of 94.1%. The increased classification rate attained by Resnet 18 can be attributed to the network's use of novel techniques to lessen over-fitting in its model.

The first method involved artificially enlarging the dataset with the aid of a label-preserving transformation. This involved extracting random patches (224x224 for ResNet 18) and training the network on them while varying the intensities of the RGB channels in the training images. The result was the generation of image translations and horizontal reflections. The second strategy was "dropout," which involves removing neurons that do not participate in the forward pass or the backward propagation. As a result, the model is forced to learn more robust characteristics and decreases the complex co-adaptations of neurons. The GPU execution time is optimal in the case of AlexNet but what requires our attention is that the pretrained models converge much faster than the CNN. The time saving is very considerable.

## 6. Conclusion

This work aimed at developing a convolutional neural network (CNN) model that will help classify COVID-19 and non-COVID 19 disease such as viral pneumonia cases using chest X-ray images in the period caused by the pandemic. The model used in this work was CNN as well as pre-trained models including AlexNet, GoogleNet, and ResNet18. The CNN gave a result with 89.13% accuracy for classifying the four classes after training 80% of the dataset and testing on 20%. This motivated us not only to keep changing settings, but also to work on pretrained model. In the latter, the pre-trained models were used on the same dataset but with just one epoch for each model. And the results were 89.93%, 91.87% and 94.1 % for AlexNet, GoogleNet and ResNet18 respectively.